\documentclass[aps,prc,twocolumn,showpacs,floatfix,nofootinbib,preprintnumbers,superscriptaddress,longbibliography,amsproc,amsmath,amssymb]{revtex4-2}

\usepackage{epsfig,dsfont,amssymb,amsmath,amsthm,amsfonts,amsbsy,mathrsfs}
\usepackage{graphicx}
\usepackage{mathtools}

\usepackage{amsmath}
\usepackage{amssymb}%
\usepackage{dcolumn}
\usepackage{hyperref}
\usepackage{soul} % does not mess up et al; use \st
\graphicspath{{./figures/}}

\usepackage[disable]{todonotes}

\newcommand{\be}{\begin{equation}}
\newcommand{\ee}{\end{equation}}
\newcommand{\ba}{\begin{eqnarray}}
\newcommand{\ea}{\end{eqnarray}}

\newcommand{\ket}[1]{\left| #1 \right\rangle}

\begin{document}

%% Suggestion for title and ordering of authors follows

\title{Scattering phase shifts from a quantum computer}

\author{Sanket Sharma}
\affiliation{Department of Physics and Astronomy, University of
  Tennessee, Knoxville, Tennessee 37996, USA}

\author{T.~Papenbrock}
\affiliation{Department of Physics and Astronomy, University of
  Tennessee, Knoxville, Tennessee 37996, USA}
\affiliation{Physics Division, Oak Ridge National Laboratory, Oak
  Ridge, Tennessee 37831, USA}

\author{L.~Platter}
\affiliation{Department of Physics and Astronomy, University of
  Tennessee, Knoxville, Tennessee 37996, USA}
\affiliation{Physics Division, Oak Ridge National Laboratory, Oak
  Ridge, Tennessee 37831, USA}

\begin{abstract}
We calculate two-body scattering phase shifts on a quantum computer using a leading order short-range effective field theory Hamiltonian. The algorithm combines the variational quantum eigensolver and the quantum subspace expansion. As an example, we consider scattering in the deuteron $^3$S$_1$ partial wave. We calculate scattering phase shifts with a quantum simulator and on real hardware. We also study how noise impacts these calculations and discuss noise mitigation required to extend our work to larger quantum processing units. With current hardware, up to five superconducting qubits can produce acceptable results, and larger calculations will require a significant noise reduction.
\end{abstract}

%\pacs{21.30.-x, 21.30.Fe, 21.10.Dr, 21.60.-n}

\maketitle

%\paragraph*{\it Introduction.---}
%\label{sec:intro}
{\it Introduction.---} Decades ago, \textcite{feynman1982} proposed quantum computers as the ultimate tool to simulate quantum mechanical systems and thought their development was a worthwhile and interesting task in itself. The recent progress in quantum hardware regarding fidelities and qubit count~\cite{byrd-ding} and in quantum algorithms has intrigued researchers across all fields of physics~\cite{cao2019,klco2022,ayral2023}. In a few years, the number of qubits used in simulations has grown from a few~\cite{kandala2017} to one hundred~\cite{farrell2023}. Nuclear physicists have embraced the potential of this new technology and explored simple models~\cite{Dumitrescu:2018njn,klco2018,cervia2021,du2021,stetcu2022,kiss2022,hlatshwayo2022,perezobiol2023}, studied entanglement~\cite{beane2019,robin2021,gu2023}, modeled neutrino physics~\cite{roggero2020,cervia2022,amitrano2023,illa2023}, and proposed algorithms for state preparations~\cite{lacroix2020,roggero2020b,choi2021,bedaque2022,coelloperez2022}.
However, the critical analysis by \textcite{Lee:2022she} 
suggests that classical computations of molecular structure in quantum chemistry might be harder to beat on quantum computers than originally thought because there are powerful classical algorithms that permit accurate computations at a cost that increases polynomially (and not exponentially) with increasing system size. In contrast, the simulation of dynamical processes in real-time still poses formidable challenges in classical computing, and many efforts are dedicated to exploring those in quantum computing~\cite{martinez2016,barison2021,klymko2022}.  

Of course, many time-dependent phenomena can be computed more efficiently in the energy domain, e.g., via the scattering matrix or response functions~\cite{efros2007,bacca2014b,roggero2018,roggero2020}. In elastic scattering, the phase shifts determine the scattering matrix, and -- strictly speaking -- their computation at arbitrary energies requires the solution of a continuum problem. However, phase shifts at discrete energies can be computed within a bound-state approach based on finite Hilbert spaces~\cite{heller1974}. Nuclear {\it ab initio} calculation of scattering phase shifts are challenging~\cite{nollett2007,quaglioni2008,shirokov2009,elhatisari2015}. 
They are also challenging for simple models on noisy hardware because energies of excited states need to be computed accurately. In this paper, we show how to meet this challenge.

One important algorithm is the variational quantum eigensolver
(VQE)~\cite{Peruzzo2014,McClean2016}. It returns the minimum energy for a 
variational ansatz provided. VQE is a hybrid algorithm
that pairs an optimizer running on a classical computer with the
evaluation of energy expectation values on a quantum device.  For
many applications in physics and chemistry, VQE has been a popular
choice since it facilitates the calculation of the ground state of an
Hamiltonian if a good trial wave function is given, see Refs.~\cite{kandala2017,omalley2016,shen2017,Dumitrescu:2018njn} for early examples and Ref.~\cite{tilly2022} for a review. 
% For example, it has been used in quantum chemistry to estimate the ground state energy of the water molecule~\cite{Nam2020} and in condensed matter physics to estimate the ground state of the Fermi-Hubbard model~\cite{Stanisic2022}. In nuclear physics, it was used to calculate the deuteron binding energy~\cite{Dumitrescu:2018njn} as well as the binding energies of $^3$H, $^3$He, and $^4$He~\cite{PhysRevA.100.012320}, and the ground state and the first excited state of $^6$Li~\cite{PhysRevC.106.034325}. 

Recently, a similar hybrid approach, known as the quantum subspace
expansion, was proposed to calculate the excited state
spectrum of a quantum mechanical system~\cite{PhysRevA.95.042308}. In this approach, one computes the Hamiltonian matrix elements 
on the quantum hardware by using excitation operators that act on the
ground state obtained from VQE. The resulting matrix is then
diagonalized on a classical computer. This approach has been used to compute the spectra of the hydrogen molecule~\cite{PhysRevX.8.011021}, to simulate spin defects~\cite{PRXQuantum.3.010339}, and for the simulation of periodic materials~\cite{PhysRevResearch.4.013052}.

Here, we extract scattering phase shifts 
using a quantum computer. Our approach combines
the VQE algorithm, finite volume methods~\cite{heller1974} developed for
the harmonic oscillator basis~\cite{shirokov2009,luu2010,More:2013rma}, and the quantum subspace expansion. At the heart of this method lies that the quantum subspace expansion yields low-lying, positive-energy eigenvalues of the Hamiltonian. In finite volume problems, these can be related to scattering observables such as phase shifts. 
We start by discussing the two-nucleon Hamiltonian and the relation between finite volume eigenvalues and phase shifts. We then present our results obtained on real hardware and simulators and discuss how noise impacts the calculations on larger QPUs. Finally, we end with a brief summary and outlook.

% \paragraph*{\bf Two-nucleon Hamiltonian -}
% \label{sec:two-nucl-hamilt}
{\it Two-nucleon Hamiltonian.---} Throughout this work we focus entirely on the $^3S_1$ partial wave of the nucleon-nucleon system. This fixes the spin/isospin degrees of freedom and we can limit the discussion to $S$ waves in the center-of-mass system. Thus, we deal with a one-body potential-scattering problem. We employ a Hamiltonian from short-range effective field theory (EFT) \cite{kaplan1998,Kaplan:1998we} that was also used in
Refs.~\cite{Dumitrescu:2018njn,shehab2019}. This EFT is a systematic low-energy expansion in powers of $R/a$, where $R$ denotes the range of the interaction and $a$ the two-body scattering length. This approach has been used successfully to calculate a number of few-nucleon observables to high accuracy \cite{Hammer:2019poc}. Here, we will use only the leading order of this approach that also maps directly onto the zero-range limit.  One possible way to implement this EFT is to use a separable interaction
\begin{equation}
V = V_0 | g \rangle \langle g|~,
\end{equation}
where $V_0$ is the coupling constant adjusted to reproduce one two-body observable for an ultraviolet cutoff $\Lambda$ that is encoded in the form factor $|g\rangle$, which we specify below. The two-body problem can be solved exactly for a separable interaction, and the $S$-wave two-body on-shell $t$-matrix becomes~\cite{Taylor:1972pty}
\begin{equation}
\label{eq:separable-t}
t(p) = \frac{4\pi}{m}\frac{\langle p| g\rangle \langle g| p\rangle}{1/V_0 -\langle g|G_0   | g\rangle}~,
\end{equation}
where $G_0(E)|q\rangle = [E-\frac{q^2}{m}+i\epsilon]^{-1}|q\rangle$ denotes the free two-body Green's function for identical particles with mass $m$.
In the finite harmonic oscillator basis with $N$ states, the Hamiltonian in the center-of-mass system of the $^3S_1$ partial wave is written as
\begin{equation}
  \label{eq:Hamiltonian}
  H_N = \sum_{n,n'=0}^{N-1}\langle n'| (T+V) | n\rangle a^\dagger_{n'} a_n~.
\end{equation}
Here, operators $a^\dagger$ ($a_n$) create (annihilate) a
two-nucleon state in the $n$th harmonic oscillator $S$-wave state. The kinetic energy and the separable potential are
\begin{align}
  \label{eq:TandV}
  \nonumber
  \langle n'|T| n\rangle  &=\frac{\hbar \omega}{2}
                            \left[
                            (2n +{\textstyle\frac{3}{2}})\delta^{n'}_n
                            +\sqrt{n(n+{\textstyle\frac{1}{2}})}\delta^{n'+1}_n\right.\\
  \nonumber
  &\left.-\sqrt{(n+1)(n+{\textstyle\frac{3}{2}})}\delta_n^{n'-1}
  \right]~,\\
  \langle n' |V| n\rangle & = V_0 \delta^0_n \delta^{n'}_{n}~,
\end{align}
where the coupling of the $V_0 =-5.68658111$~MeV is adjusted to
reproduce the deuteron binding energy in the limit of an infinite number of harmonic oscillator
states and $\hbar\omega=7$~MeV. The Kronecker $\delta$ functions reflect that we use the lowest harmonic oscillator orbital for the form factor $|g\rangle$, i.e. 
$\langle q|g \rangle = b^{1/2}\pi^{-1/4} \exp{(-q^2 b^2/2)}$,
where $b$ is the oscillator length.

Though we are dealing technically with a one-body system, the second-quantized form of the Hamiltonian~(\ref{eq:Hamiltonian}) is attractive  for quantum computing. The operators $a^\dagger_n$ and $a_n$ become spin-lowering and spin-rising operators, respectively, when acting on qubits, and the mapping  
uses the Jordan-Wigner transform~\cite{Dumitrescu:2018njn}. Within this work, an unoccupied and occupied state $|n\rangle$ of the harmonic oscillator will correspond to the state $\ket{0}$ and $\ket{1}$, respectively, of qubit $n$.
% \begin{align}
%   \label{eq:J-W transform}
%   a^\dagger_{n} \xrightarrow{} \frac{1}{2} \left[\prod_{j=0}^{n-1} -Z_{j} 
%   \right] \left(X_n - iY_n\right) \\
%   a_{n} \xrightarrow{} \frac{1}{2} \left[\prod_{j=0}^{n-1} -Z_{j} 
%   \right] \left(X_n + iY_n\right)
% \end{align}
% [[LP: Remove?]]
% The Hamiltonian above can be implemented on a quantum computer by
% interpreting the $|0, \ldots, 0,1,0,\ldots \rangle $ as the $i$th harmonic-oscillator
% basis state. 
We use the VQE algorithm to find the ground state of the system. This requires an ansatz with parameters that can be optimized to find the lowest possible energy. We use the unitary coupled cluster (UCC) ansatz also used in Refs.~\cite{Dumitrescu:2018njn,shehab2019}. The ansatz circuit we implemented is similar to the one in~\cite{shehab2019}.

% \paragraph*{\bf The quantum subspace expansion -}
% \label{sec:quant-subsp-expans}
{\it The quantum subspace expansion.---} The quantum subspace expansion was developed by \textcite{PhysRevA.95.042308}
to calculate excited states. Like VQE, it is a hybrid algorithm that
relies on a combination of classical and quantum computing. First, the
VQE algorithm is used to generate the ground state wave function of
the Hamiltonian under consideration. This ground-state wave function
$|\Psi\rangle$ is used in combination with excitation operators to
generate a subspace in which the Hamiltonian can be diagonalized with
a classical computer. Here, we will use single-particle excitation operators
%\begin{equation}
%  \label{eq:excitation}
%  E = \{e_{j l} \,|\, j,l \, \epsilon \, 0, 1...\}
%\end{equation}
%where
\begin{equation}
   \label{eq:creationannihilation}
     e_{\alpha} = a^\dagger_j a_l~.
\end{equation}
where $j, l = 0, \ldots N-1$ and $\alpha$ is a single label that uniquely identifies $j$ and $l$ (e.g. $\alpha = Nj+l)$). We note that this approach identifies all excitations in the Hilbert space. This completeness of the basis is not achievable in many practical applications due to the enormous size of Hilbert spaces. Instead, one could identify a subset of relevant excitations and then diagonalize the Hamiltonian in that basis. This is essentially the generator coordinate method~\cite{ringschuck}.  
% \begin{equation}
%     \label{eq:creationannihilation}
%       e_{\alpha} = a^\dagger_{\alpha} a_0~.
% \end{equation}
% where $\alpha = 0, \ldots N-1$.   

On the quantum computer, we evaluate the matrix elements 
\begin{equation}
  \label{eq:subspacematrix}
  \Tilde{H}_{\alpha\beta}= \langle \Psi| e^\dagger_\alpha H e_\beta |\Psi\rangle~,
\end{equation}
and the overlap matrix elements
\begin{equation}
  \label{eq:overlapmatrix}
  \Tilde{S}_{\alpha\beta}= \langle \Psi| e^\dagger_\alpha e_\beta |\Psi\rangle~.
\end{equation}
Now, we solve the generalized eigenvalue problem
\begin{equation}
    \Tilde{H}\ket{\Psi} = E \Tilde{S}\ket{\Psi}
\end{equation}
on a classical computer and employ the usual techniques, see, e.g., Ref.~\cite{ringschuck}.
% We convert it into an ordinary eigenvalue problem using the below approach.
% \begin{equation}
%   \label{eq:generaleigenvalue}
%   \underbrace{S^{-1/2}HS^{-1/2}}_{\Tilde{H}}\underbrace{S^{1/2}\ket{\Psi}}_{\ket{\Phi}} = E'\underbrace{S^{1/2}\ket{\Psi}}_{\ket{\Phi}}     
% \end{equation}

% To compute $S^{-1/2}$ we diagonalize S such that
% \begin{align}
%     S &= U^\dagger S' U \\
%     S^{-1/2} &= U^\dagger S'^{-1/2}U
% \end{align}

% We discard eigenvectors corresponding to those eigenvalues to deal with the scenario when one or more eigenvalue(s) are zero or very close to zero.
% Now,
% \begin{equation}
%     \Tilde{S}^{-1/2} = \Tilde{U}^{\dagger}\Tilde{S'}^{-1/2}\Tilde{U}
% \end{equation}
% where $\Tilde{S'}$ is the diagonal matrix obtained after removing zero eigenvalues and $\Tilde{U}$ is the matrix of eigenvectors corresponding to those eigenvalues.
% \begin{equation}
%     \Tilde{H} = \Tilde{S}^{-1/2} H \Tilde{S}^{-1/2}
% \end{equation}
% Finally, we solve the eigenvalues problem 
% \begin{equation}
%     \Tilde{H} \ket{\Phi} = E \ket{\Phi},
% \end{equation}
% to obtain the energies of the excited states $E'_i$ and the corresponding momenta $k_i=\sqrt{2\mu E_i'}$.
This overlap matrix has only $N$ non-zero eigenvalues, and we discard the eigenvectors corresponding to zero eigenvalues. In the presence of noise, we keep the $N$ eigenvectors with the largest eigenvalues. The resulting energy eigenvalues are $E_i$, and the corresponding momenta are denoted as $k_i=\sqrt{2\mu E_i}$. 
All our QSE results presented in this work were obtained as described above. However, this $N^2\times N^2$ problem can be reduced to an $N\times N$ problem by setting $l$ to 0 in Eq.~\eqref{eq:creationannihilation} and solving the resulting smaller generalized eigenvalue problem. We have verified on the simulator for the five- and seven-qubit systems that this approach leads to the same results within statistical fluctuations. 
%The resulting energy eigenvalues are $E_i$, and the corresponding momenta are denoted as $k_i=\sqrt{2\mu E_i}$.

% \begin{figure}[!htbp]
%     \includegraphics[width=0.99\linewidth]{figure_name.pdf}
%     \caption{Figure caption.}
%     \label{fig:snm}
% \end{figure}

\begin{figure}
    \includegraphics{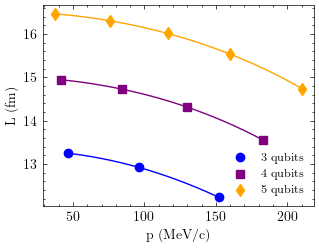}
    \caption{Box size $L$ as a function of the relative momentum $p$ for the simulations of systems with 3, 4, and 5 qubits. The markers are at the relative momenta of the eigenvalues of the kinetic energy. The solid lines are interpolation curves.}
    \label{fig:L-vs-p}
\end{figure}

Having obtained the ground state wave function $|\Psi\rangle$ on the quantum device, we then obtain the phase shifts from the positive energy eigenvalues of the
Hamiltonian. For this, we follow the approach described in Ref.~\cite{More:2013rma}. The key insight is briefly described as follows:
Employing a finite harmonic oscillator basis transforms the scattering
continuum into a discrete set of states confined within a finite
volume. This effectively imposes Dirichlet boundary conditions at
a specific radial distance $r = L$.
The hard wall radius $L$ depends on $N$, $\hbar\omega$, and the scattering energy. To compute $L$, one must first diagonalize the kinetic energy operator $\hat{T}$ in the finite harmonic oscillator basis. This yields a spectrum of energies $T_i$ and corresponding momenta $p_i = \sqrt{2 \mu T_i}$. In position space, the corresponding eigenfunctions resemble those of a spherical cavity with a radius $L$. The location of the $i^{\rm th}$ zero of the spherical $S$-wave Bessel function $j_0(p_i L_i\hbar)$ then determines $L=L(p_i)\equiv L_i$. From the resulting set of $p_i$ and $L_i$ we can then generate a smooth and continuous interpolation function $L(p)$. Figure \ref{fig:L-vs-p} shows box sizes $L(p)$ as a function of the relative momentum $p$ for 3-, 4-, and 5-qubit systems. Given that the kinetic energy operator $\hat{T}$ is diagonalized classically, the calculation of $L(p)$ in unaffected by quantum noise. The observed convexity and small curvature of the resulting curves suggest that these interpolations are very reliable, leaving therefore very little room for additional uncertainties due to this approach of determining the effective box size $L(\hbar k_i)$.

The phase shift for the $i$th momentum $k_i$ is then given by
\begin{equation}
  \label{eq:phase-shift}
  \tan \delta_0(k_i) = \frac{j_0 (k_i L(\hbar k_i))}{\eta_0( k_i L(\hbar k_i))}~,
\end{equation}
where $j_0$ and $\eta_0$ denote the $S$-wave spherical Bessel and Neumann functions, respectively. The above equation reflects that the radial wave function has to be zero at $L$ as required by the hard wall boundary condition. When $L$ is known, Eq.~(\ref{eq:phase-shift}) yields a result that matches exactly with the analytical result.

% \paragraph*{\bf Results -}
{\it Results.---}
\begin{figure}[t ]
\begin{center}
     \includegraphics[width=0.9\linewidth]{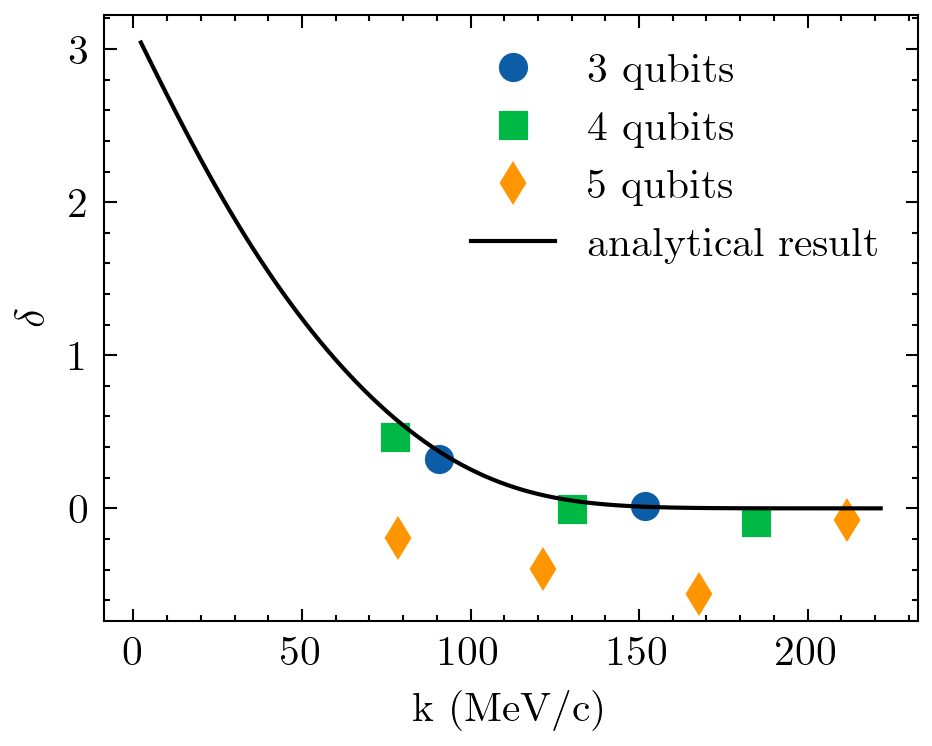}
\end{center}     
     \caption{\label{fig:snm} (Color online) Scattering phase shifts $\delta$ as a function of the relative momentum $k$. The solid line denotes the analytical results. The circles, squares, and diamonds give the results obtained with a 3-, 4-, and 5-qubit simulation, respectively. Simulation results were obtained on the \textit{ibmq\_jakarta} QPU.}
\end{figure}
Our quantum computations used IBMQ machines.  Throughout this work, we employed noise mitigation of readout errors. We measured the complete assignment matrix to perform readout error mitigation on the noisy counts received from the QPU. For this, we include circuits in which we initialize the qubits to combinations of zeros and ones. We then multiply the inverse of this assignment matrix with the counts to obtain the noise-mitigated counts. 
% In our calculations involving three qubits, we calculate the ground state by dividing the range of two parameters into a grid and then brute forcing over it by calculating the expectation value of the Hamiltonian for each point on the grid. 
For three qubits, the ground state depends on two parameters, and we compute energy expectation values on the quantum hardware using a two-dimensional grid. In our calculations involving more than three qubits, we assume that the deuteron ground state was previously determined by techniques such as VQE and instead use the 
ground state from an exact diagonalization.
% and will not address this part of the calculation. 
We pass the exact ground state to our code and then evaluate the matrix elements required for the quantum subspace expansion on quantum hardware. The computations on $N$ qubits yield $N-1$ excited states, and we determine the corresponding phase shifts using Eq.~(\ref{eq:phase-shift}).

In Fig.~\ref{fig:snm}, we show results for three, four, and five qubits on the \textit{ibmq\_jakarta} QPU as circles, squares, and diamonds, respectively. The solid line shows the results obtained from the analytical expression given in Eq.~\eqref{eq:separable-t}. Our results for the three qubits agree with the analytical result. This is encouraging as the corresponding bound state calculation requires error mitigation to obtain accurate results~\cite{Dumitrescu:2018njn}. As we increase the number of qubits, the agreement worsens, and the five qubit results are inaccurate.

Our method produces phase shifts at momenta determined by the harmonic oscillator frequency. The effective field theory allows us to vary the harmonic oscillator frequency (while adjusting the potential strength $V_0$ such that the deuteron bound state energy is reproduced correctly in the infinite oscillator basis). This permits us to compute phase shifts at different momenta at a fixed number of qubits $N$. 

\begin{figure}[t]
\begin{center}
\includegraphics[width=0.9\linewidth]{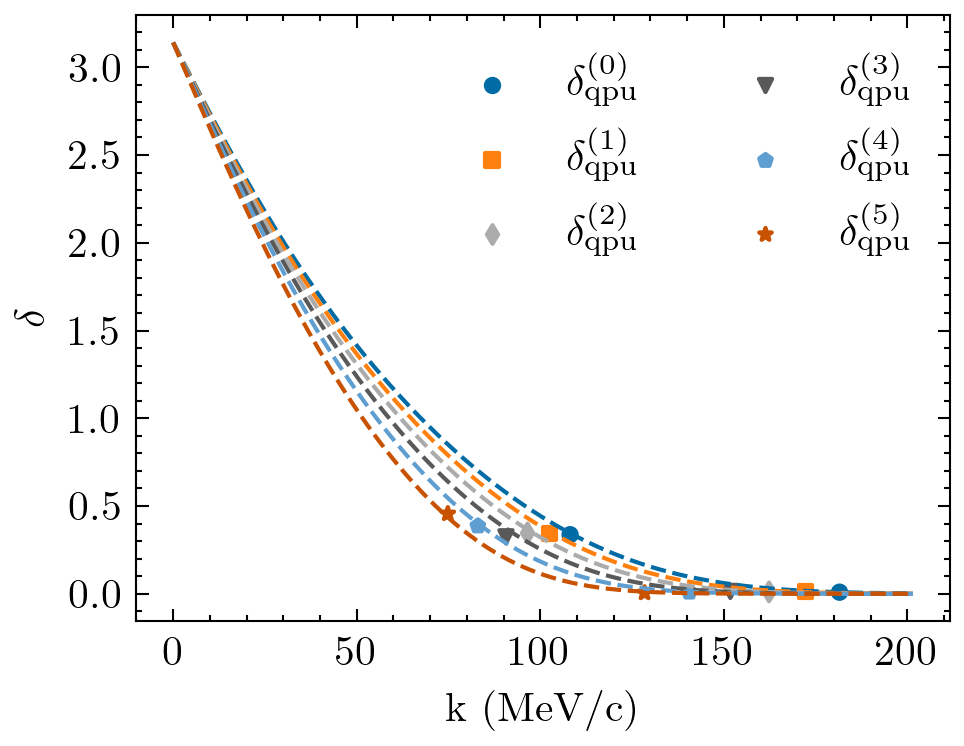}
\end{center}   
\caption{\label{fig:tapj} (Color online) The data points are the phase shifts obtained using six different interactions (with the oscillator frequency and the momentum cutoff increasing from left to right) using three qubits. 
The dashed lines are the analytical results corresponding to the interaction represented by the same color. The circles denote the results obtained with the QPU \textit{ibmq\_jakarta}.}
\end{figure}

We generated five additional interactions and now consider oscillator frequencies $\hbar\omega =5, 6, 7, 8, 9, 10$~MeV. 
The corresponding momentum cutoffs are $\Lambda= 128, 140, 152, 162, 172, 181$~MeV, respectively. For each frequency, we repeat the calculations of phase shifts and show the corresponding results, obtained on three qubits, in Fig.~\ref{fig:tapj}. The analytical results for our six interactions vary slightly due to remaining regulator dependence in the renormalization process: All interactions essentially reproduce the scattering length (i.e., they agree on the slope of the phase shifts at zero momentum) but differ in the effective range. We see that the analytical results differ, as expected, at larger momenta and that the phase shifts vanish at momenta above the respective cutoff. In all cases, the results from quantum computing agree with the corresponding analytical data.

\begin{figure}[t]
\begin{center}
\includegraphics[width=0.9\linewidth]{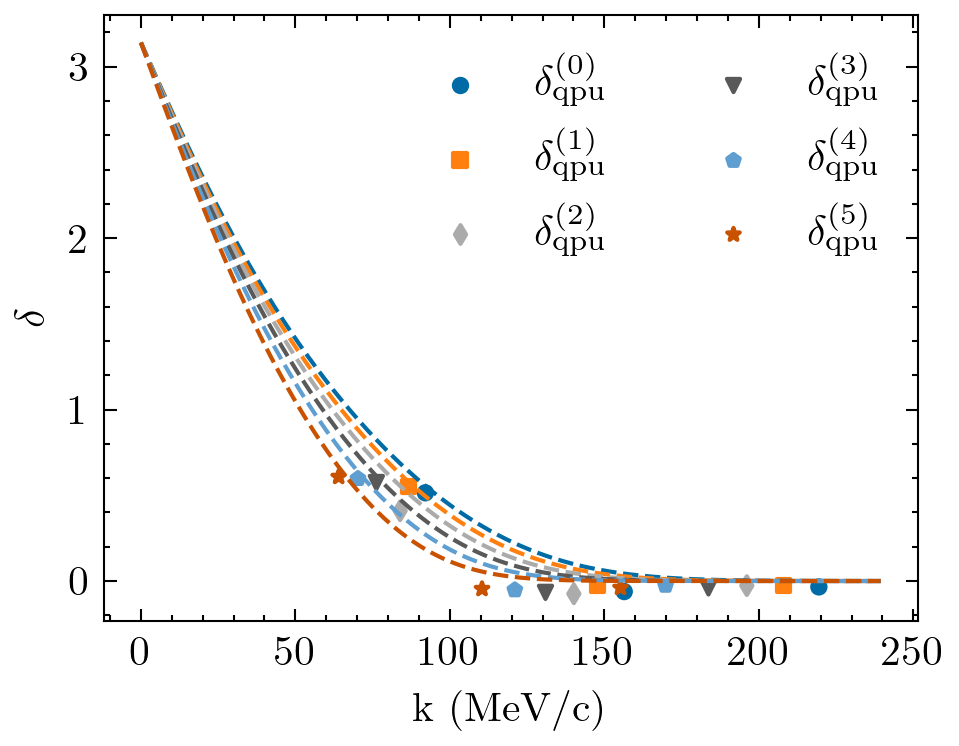}
\end{center}   
\caption{\label{fig:fapn} (Color online) Phase shifts obtained using six different interactions using four qubits.  The dashed lines denote the analytical results corresponding to the interaction represented by the same color. The circles denote the results obtained with the QPU \textit{ibm\_nairobi}.}
\end{figure}

Figure~\ref{fig:fapn} shows the results obtained on four qubits, using the QPU \textit{ibm\_nairobi}. Here, the low-energy phase shifts are accurate. At higher energies, the phase shifts are still close to the analytical results, but they essentially vanish. This suggests that it will be challenging to employ more qubits in the computations using quantum hardware.

\begin{figure}[htb]
\begin{center}
\includegraphics[width=0.9\linewidth]{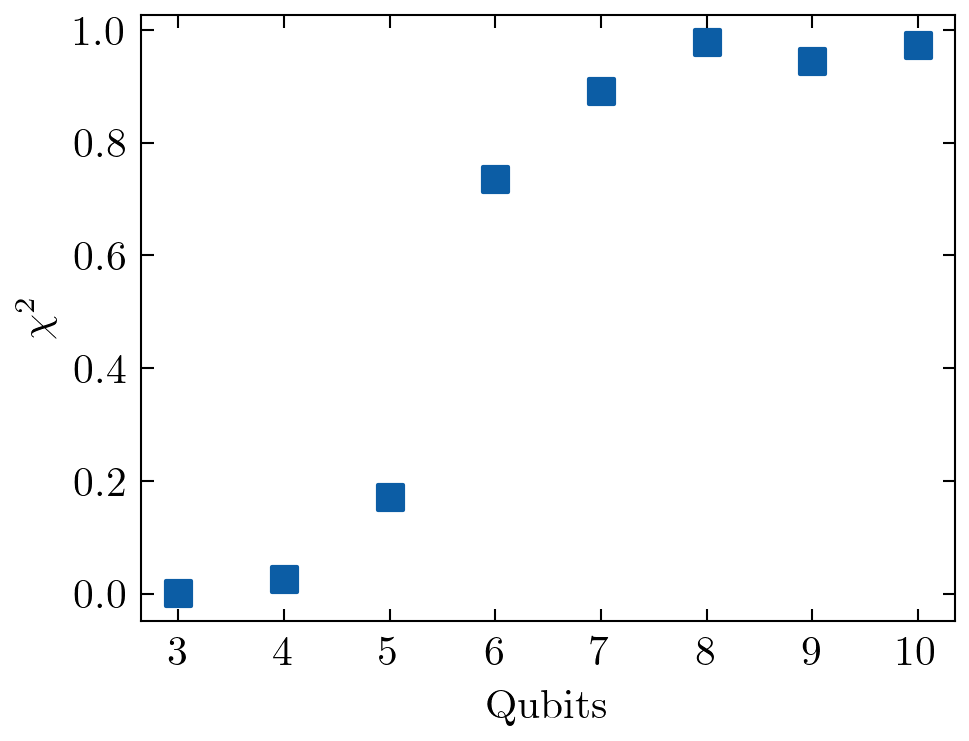}
\end{center}   
\caption{\label{fig:csfn} (Color online) $\chi^2$ as a function of number of qubits}
\end{figure}

We computed the total $\chi^2$ deviation between the phase shifts from simulations 
%[Lucas: we should mention that we used the $\textit{ibmq\_guadalupe}$ noise model for these simulations here and also in the captions of the two plots]
and analytical results based on a model space of $N$ qubits. The simulations employ the $\textit{ibmq\_guadalupe}$ noise model.  The results, shown in Fig.~\ref{fig:csfn}, suggest that there is a transition at about $N\approx 5$ beyond which quantum computations become too limited by noise.    

\begin{figure}[t]
\begin{center}
     \includegraphics[width=0.9\linewidth]{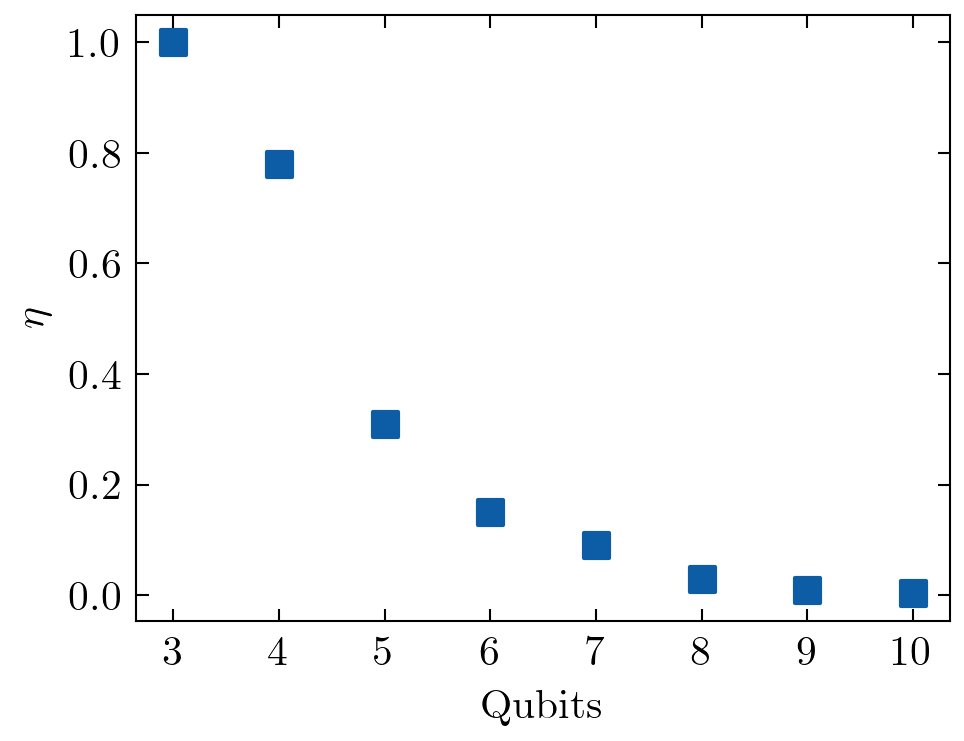}
\end{center}     
     \caption{\label{fig:evq} Estimated noise scaling factor $\eta$ required to produce acceptable results as a function of the number of qubits.}
\end{figure}

To understand this apparent limitation, we studied the impact of noise on larger systems using a QPU simulator using again the $\textit{ibmq\_guadalupe}$ noise model. We are interested in quantifying how much noise is tolerable to compute accurate phase shifts on a fixed number of qubits. To address this point, we modified Qiskit's noise model source code. We introduced a single scaling factor $\eta\le 1$ that simultaneously reduces the one-qubit gate errors, two-qubit gate errors, and the readout error. For a given system size $N$ we decreased $\eta$ using the \textit{ibmq\_guadalupe} noise model until an approximate agreement between simulation and analytical results was achieved. The estimated value of $\eta$ obtained in this way is plotted against the number of qubits in Fig.~\ref{fig:evq}. We see that there is a jump as one goes from four to five qubits, and an order-of-magnitude in noise reduction is required for computations on more than six qubits.

% \paragraph*{\bf Summary -}
%\label{sec:summary}
{\it Summary.---} We computed nucleon-nucleon scattering phase shifts on a quantum device by combining hybrid quantum algorithms with finite-volume approaches to scattering. We employed the quantum subspace expansion to compute discrete continuum states and mapped those to scattering phase shifts. While current superconducting hardware allows one only to treat simple models, {\it ab initio} computations of scattering phase shifts are expensive~\cite{nollett2007,quaglioni2008,shirokov2009,elhatisari2015}. It is here that error-corrected quantum computing could be advantageous in the future.      

% We have performed the first calculation of scattering phase shifts for the two-nucleon system on quantum simulators and real hardware with the quantum subspace expansion. On the one hand, we used that the truncated harmonic oscillator basis maps the scattering problem on a finite volume problem, and on the other, that the quantum subspace expansion facilitates a calculation of higher lying eigenvalues of the resulting finite volume Hamiltonian.

% The momenta at which phase shifts are calculated depends on the parameters of the Hamiltonian. When an effective theory describes the physical system, the freedom to change the regulator scale can be exploited to modify the momentum range.

Using noise mitigation to correct readout errors only allowed us to perform accurate computations with up to four or five superconducting qubits. Our study of this problem revealed that an order-of-magnitude reduction in readout errors and one- and two-qubit gate errors is necessary for accurate quantum computations with significantly more qubits.  We believe that this result will also hold for other applications that can be qualified as {\it dense} problems in which all qubits need to be entangled.
In the future, it would be interesting to analyze whether additional noise mitigation techniques, such as Richardson extrapolation or randomized compiling, can move this boundary significantly upwards.

\begin{acknowledgments}
We acknowledge valuable discussions with Titus Morris.
This work was supported by the U.S. Department of Energy, Office of Science, Office of Nuclear Physics, under Award
Nos.~DE-SC0021642 and DE-FG02-96ER40963, by the National Science Foundation under Grant No. PHY-2111426, and by the Quantum Science Center, a National Quantum Information Science Research Center of the U.S. Department of Energy. Oak Ridge National Laboratory is supported by the Office of Science of the U.S. Department of Energy under Contract No. DE-AC05-00OR22725.  Access to the IBM Quantum Services were obtained through the IBM Quantum Hub at Oak Ridge. The views expressed are those of the authors and do not reflect the official policy or position of IBM or the IBM~Q team.
\end{acknowledgments}

%\bibliography{qcqis,master}


\end{document}